# Directly determining orbital angular momentum of ultrashort Laguerre–Gauss pulses via autocorrelation measurement

Bing-Shi Yu[1], Chun-Yu Li[1], Yuanjie Yang[2], Carmelo Rosales-Guzmán[1,3], and Zhi-Han Zhu[1]

[1] Wang Da-Heng Center, HLJ Key Laboratory of Quantum Control, Harbin University of Science and Technology, Harbin, China
[2] School of Physics, University of Electronic Science and Technology of China, Chengdu, China
[3] Centro de Investigaciones en Óptica, A.C., Loma del Bosque 115, Colonia Lomas del campestre, 37150 León, Gto., Mexico

E-mail: dr.yang2003@uestc.edu.cn, carmelorosalesg@cio.mx, and zhuzhihan@hrbust.edu.cn

**Abstract**

Autocorrelation measurement based on second-harmonic generation (SHG), the best-known technique for measuring the temporal duration of ultrashort pulses, could date back to the birth of ultrafast lasers. Here, we propose and experimentally demonstrate that such well-established technique can also be used to measure the orbital angular momentum of ultrashort Laguerre–Gauss (LG) pulses. By analysing the far-field pattern of the SHG signal, the full spatial structure of ultrashort LG pulses, including both azimuthal and radial indices, are unambiguously determined. Our results provide an important advancement for the well-established autocorrelation technique by extending it to reach its full potential in laser characterization, especially for structured ultrashort pulses.

**Introduction**

Ultrashort pulses carrying orbital angular momentum (OAM) have gained an increasing interest in the community of ultrafast photonics and structured light [1-9]. An important reason for this is that optical vortices with ultrahigh peak power and ultrashort duration provide a new tool to explore novel applications, as well as fundamental aspects of physics. For instance, ultrashort vortex pulses structured in their spatial or polarisation degrees of freedom, can produce light fields with extraordinary spatiotemporal and topological structures [10-14]. In a similar way, high-order harmonic generation of ultrashort vortex pulses can provide a wide variety of novel light-matter interactions [15-19]. It is worth mentioning that most of the ultrashort vortex pulses in the aforementioned studies focus only on Hypergeometric-Gaussian vortex modes embedded with OAM, which are not eigenmodes of the paraxial-wave equations (PWE) with propagation-invariant transverse structure in free space [20]. For a further study it is required to consider OAM modes with well-defined radial structure, such as Laguerre-Gaussian (LG) modes, which are indeed solutions of the PWE in cylindrical coordinates [21]. Importantly, such propagation-invariant modes play an important role in various applications [22,23]. It is therefore not surprising that the generation and applications of ultrashort pulsed Laguerre–Gaussian (LG) modes have aroused intense interest recently [2, 24]. As such, a proper characterization of these modes is a curial toolkit in various research fields.

In this way, full characterization of LG modes requires to determine simultaneously their azimuthal and radial indices. A common way to accomplish this task involves mode-projective measurement, which require the projection of the unknown LG mode onto several elements of the LG basis, which is usually implemented through spatial light modulators [25,26]. Importantly, this commonly used technique requires precise *a priori* knowledge of some of the parameters of the unknown beam, such as, its initial beam width radius $w_0$ and its radius of curvature $R_z$, for an accurate characterisation of the mode in question [27-29]. For ultrashort pulses, transverse dispersion induced by the grating of complex-amplitude modulation would bring additional issues. Although a recent work has suggested that one can measure OAM of intense ultrashort pulses by exploiting strong-field ionization [30], a simple technique towards commercial use is still to be developed.

More recently, the idea of the mode-projective measurement was used with nonlinear interactions and applied to numerous novel concepts, such as nonlinear modal detection and spatial mode teleportation [31,32], enriching the already active topic of structured nonlinear optics [30-40]. The history of ultrashort-pulse characterization with nonlinear optics dates back to the birth of ultrafast lasers. Perhaps one of the most known techniques is the so-called autocorrelation technique based on second-harmonic generation (SHG) that was widely used to measure the temporal property of ultrashort pulses [41]. In this work, we extend this well-established technique to characterize the spatial structure of ultrashort pulses. By theoretical analysis and experimental verification, we show that the far-field pattern of the SHG is equivalent to the autocorrelation function of spatial modes and thus, in addition to measuring its pulse width, it can also be used to determine the OAM content of ultrashort pulsed LG modes.

## Methods and Results

Laguerre-Gaussian modes represent a well-known family of paraxial beams featuring a ring-shaped (or donut-like) intensity pattern, except for those with a topological charge $\ell = 0$. As eigensolutions of the PWE, any LG mode, or modes resulting from superpositions of the same modal order $N$, maintain their transverse pattern upon propagation (termed self-similar beams), apart from a diameter scaling, $w_z = w_0[1+(2z/kw_0^2)^2]^{1/2}$, where $w_0$ is the beam radius at the waist plane. The spatial complex amplitude of an LG mode in cylindrical coordinates $\{r,\varphi,z\}$ has the form,

$$LG_p^\ell(r,\varphi,z) = \sqrt{\frac{2p!}{\pi(p+|\ell|)!}} \frac{1}{w_z} \left(\frac{\sqrt{2}r}{w_z}\right)^{|\ell|} L_p^{|\ell|}\left(\frac{2r^2}{w_z^2}\right) \exp\left(\frac{-r^2}{w_z^2}\right)$$
$$\times \exp\left\{-i\left[kz + \frac{kr^2}{2R_z} + \ell\varphi - (N+1)\tan^{-1}\left(\frac{z}{z_R}\right)\right]\right\}, \tag{1}$$

where $\ell = 0, \pm1, \pm2, ...$ and $p = 0, 1, 2, ...$ are the azimuthal (topological charge) and radial indexes, respectively. The function $L_p^{|\ell|}(\bullet)$ is the associated Laguerre polynomial, which determines the radial amplitude structure of the mode in the transverse plane, and $N = 2p + |\ell|$ defines its modal order. The parameters $z_R = kw_0^2/2$ and $R_z = z^2 + z_R^2/z$ are the Rayleigh length and radius of curvature, respectively. The factor $\exp(i\ell\varphi)$ in Eq. (1) is an eigenfunction of the operator $-i\partial_\varphi = -i(x\partial_y - y\partial_y)$; that is, LG modes with $\ell \neq 0$ are natural carriers of optical OAM and the associated eigenvalue $\ell$ is the OAM quantum number. Moreover, because optical OAM originates from twisted wavefronts involving phase singularities, the azimuthal index is also known as "topological charge".

To fully characterize an unknown LG mode, we must determine its azimuthal ($\ell$) and radial (p) indices simultaneously. Noteworthy, we can determine the latter directly by counting the number of rings or dark nodes in the ring-shaped pattern of an LG mode, as shown in the examples in Fig. 1(c). Therefore, the main task is to determine the azimuthal index (or topological charge $\ell$) of an unknown LG mode through an autocorrelation measurement based on SHG, which is usually used to characterize the temporal duration of ultrashort pulses. To study the autocorrelation in the spatial degree of freedom, we consider the SHG of two conjugated LG modes whose spatial complex amplitude at the generation plane (usually inside the crystal) is given by

$$E^{2\omega}(r,\varphi,z_0) = LG_p^{+\ell}(r,\varphi,z_0)LG_p^{-\ell}(r,\varphi,z_0)$$
$$= u^{2\omega}(r,z_0)L_p^{|+\ell|}\left(\frac{2r^2}{w_z^2}\right)L_p^{|-\ell|}\left(\frac{2r^2}{w_z^2}\right), \tag{2}$$

where $u^{2\omega}(r,z_0)$ is the amplitude envelope of SHG at the generation plane. Hence, the spatial complex amplitude at the far field is derived through the Fourier transformation of Eq. (2), namely,

$$E^{2\omega}(r,\varphi,z_\infty) = \mathcal{F}\left[E^{2\omega}(r,\varphi,z_0)\right]$$
$$= u^{2\omega}(r,z_\infty)\rho(\zeta), \tag{3}$$

where $u^{2\omega}(r,z_\infty)$ is the amplitude envelope at the Fourier plane (far field), $\zeta$ is a real function associated with scaling factors of the Fourier lens, and $\rho(\zeta)$ governs the far-field radial structure that can be further factorized as (see Ref. 37 for more details)

$$\rho(\zeta) = \frac{(p+|\ell|)!}{p!} L_{p+|\ell|}^0(\zeta) L_p^0(\zeta), \quad \text{for } \ell \neq 0, \tag{4}$$

$$\rho(\zeta) = L_p^0(\zeta) L_p^0(\zeta), \quad \text{for } \ell = 0. \tag{5}$$

We first consider the general case $\ell \neq 0$, which is given in Eq. (4). Crucially, here the product $L_{p+|\ell|}^0(\bullet)L_p^0(\bullet)$ contains $n = 2p + |\ell| = N$ zeros, as shows the example in Fig. 1(a), which means there are $N$ phase jumps in the far-field wavefront along the radial direction. Due to the zero amplitude

at the phase jumps, $n = N$ dark nodes occur in the far-field pattern of the SHG. Hence, we can determine the azimuthal index of a LG mode directly from its far-field autocorrelation pattern, that is, by using the relation $|\ell| = N - 2p$, where the index $p$ is known directly from the dark-node number of fundamental-frequency pattern. For the second special case, we see that the term $L_p^0(\bullet)L_p^0(\bullet)$ in Eq. (5) is the same as that in Eq. (2) when $\ell = 0$. Thus, the radial structure of the SHG beam in the far field is the same as that in the generation plane, in which $\rho(\zeta)$ has $n = p$ zeros, as shown in the example of Fig. 2(b). In other words, the pattern of SHG in this special case is just the square of the input LG-mode pattern and undergoes self-imaging from the near field ($z_0$) to the far field ($z_\infty$) [33,40], see Appendix B for details.

Overall, the number of dark nodes ($n$) in the autocorrelation patterns, which can be used to determine the LG mode in both cases, is given as

$$\begin{cases} n = 2p + |\ell| & \text{for } \ell \neq 0 \\ n = p & \text{for } \ell = 0. \end{cases} \quad (6)$$

To exemplify this, Fig. 1(c) shows the spatial complex amplitude of LG modes for different azimuthal and radial indices, and Fig. 1(d) shows their corresponding spatial autocorrelation, i.e., their SHG patterns at the far field. Here, to clearly show the dark nodes, the beam profiles in Fig. 1(d) are drawn with saturated intensity. Notice that, (i) for the common case $\ell \neq 0$, one can directly measure the radial index $p$ and modal order $N$ of LG modes from the dark-node numbers in their beam intensity patterns ($p$) and associated autocorrelation patterns ($n$), as shown in Figs. 1(c) and (d), respectively, and thus can further determine their topological charges via the relation $|\ell| = n - 2p$. (ii) For the special case $\ell = 0$, we recognize the radial index $p$ from the dark-node numbers in their autocorrelations patterns, which actually are the same as the dark-node numbers in the original beam profiles shown in Fig. 1 (i.e., $n = p$).

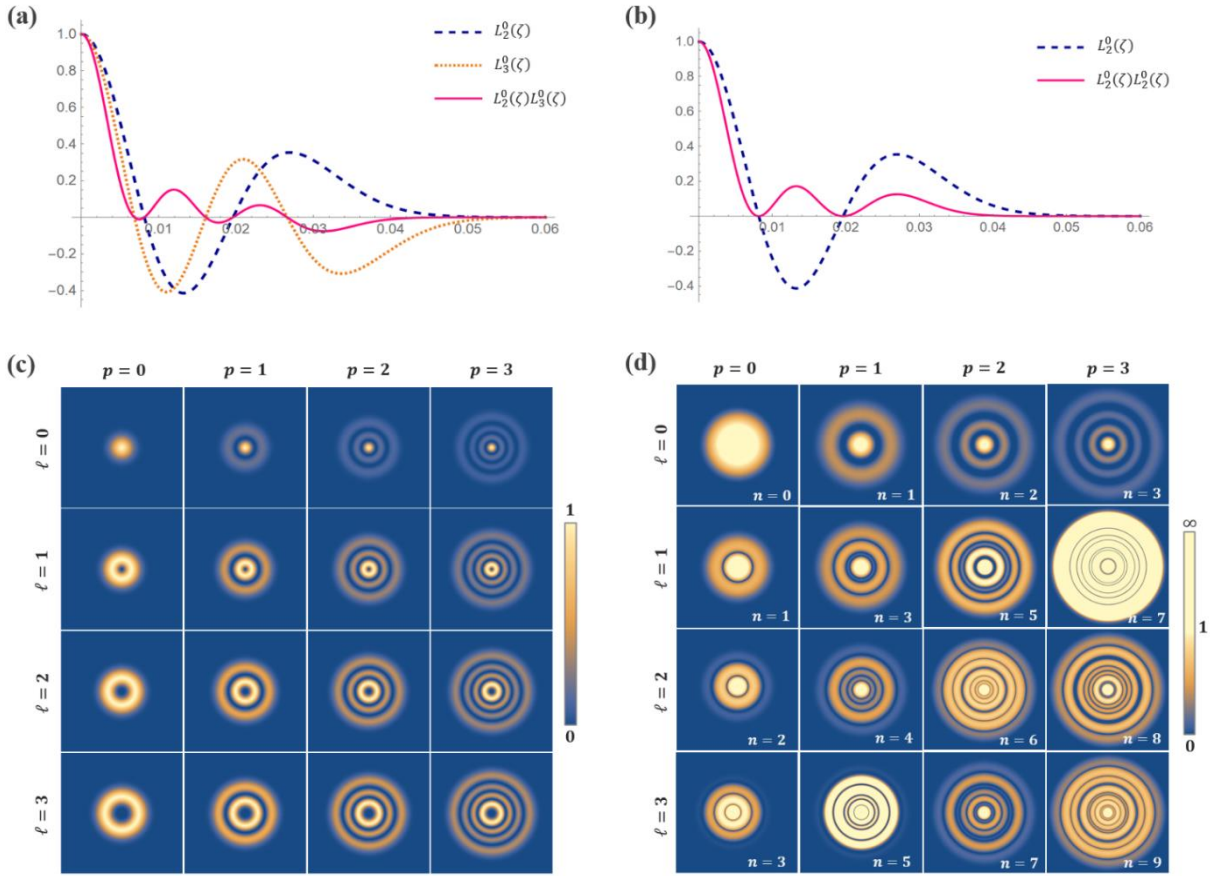

Figure 1. Theoretical Results: (a) and (b) are simulated $\rho(\zeta)$ curves for autocorrelations of $LG_2^1$ and $LG_2^0$, respectively. (c) and (d) are simulated beam profiles of LG modes and their corresponding spatial autocorrelations, respectively, where patterns in (d) are drawn with saturated intensity. The phase profiles of (c) and (d) are given in Appendix A.

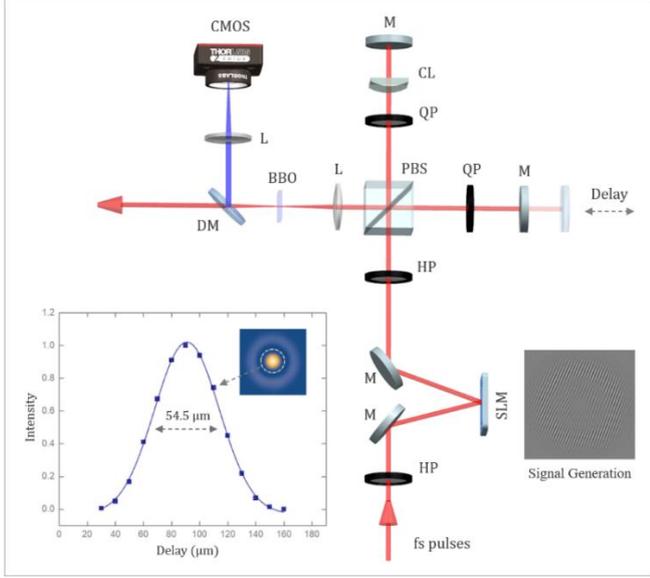

Figure 2. Schematic diagram of the implemented experimental setup, where the key components are the mirror (M), lens (L), cylindrical lens (CL), polarizing beam splitter (PBS), half-wave plate (HP), quarter-wave plate (QP), spatial light modulator (SLM), dichroic mirror (DM), and camera (CMOS).

In the following, we verify the above theory experimentally. Figure 2 shows a schematic diagram of the experimental setup, in which we use a collinear autocorrelation framework based on type-II SHG. More specifically, the signal source was an ultrafast laser operating at a wavelength of 800 nm and with a repetition rate of 80 MHz. The input laser beam was first prepared into the desired LG mode by using a complex-amplitude hologram displayed on a spatial light modulator. The prepared LG mode was then set to a diagonally polarized state and sent afterwards to a polarizing Michelson interferometer. In one arm, a mirror mounted on a translation stage was used to control the pulse delay and, in the other arm, a cylindrical lens was used to conjugate the OAM of the input LG mode. As a result, the output beam from the interferometer is a *partial* cylindrical vector beam, i.e., $1/\sqrt{2}[LG_p^\ell(t)\hat{e}_H + LG_p^{-\ell}(t-\tau)\hat{e}_v]$. Here the term 'partial' means that the cylindrical vector mode is not necessarily in a pure vector state due to the delay $\tau$. The output cylindrical vector mode was then focused into a 0.5 mm $\beta$-BaB$_2$O$_4$ (BBO) crystal with type-II phase matching to generate the SHG (i.e., autocorrelation signal). Finally, the autocorrelation signal was focused by a Fourier lens and recorded by a CMOS-based camera.

The inset at the right-bottom of Fig. 2 shows the measured temporal autocorrelation signal i.e., $I^{2\omega}(\tau) = \int I^\omega(t)I^\omega(t-\tau)dt$, where a 54.5 μm full width at half maximum (FWHM) indicates that the temporal duration of the input pulses is approximately 128 fs. Regarding this collinear SHG configuration, it is important to note that, even if the polarizing extinction ratio of the type-II SHG is not good enough, we can still obtain a great signal to noise ratio. Because the temporal signal was obtained by adding the intensities of the pixels in the central region (i.e., the Gaussian beam surrounded by dashed ring) of the SHG pattern. Afterwards, the path delay was modified to obtain the maximum intensity of the temporal signal, the spatial autocorrelation signal (i.e., the far-field SHG pattern) is recorded with a camera. Figures 3(a) and 3(b) show the measured LG modes with different modal indexes ($\ell$ and $p$) and their corresponding spatial autocorrelations, respectively. These results show that, for the common case $\ell \neq 0$, the dark-node number in the far-field SHG pattern is exactly the same as the measured modal order of LG modes (i.e., $n = 2p + |\ell| = N$). For the special case, the observed autocorrelation patterns of LG modes with $\ell = 0$ are exactly the same as the square of their beam profiles shown in Fig. 3(a) (i.e., $n = p$). In addition, since we used a low-cost CMOS with only an 8-bit dynamic range, all the autocorrelation patterns shown in Fig. 3(b) are overexposure to see the dark-node numbers clearly.

The experiment verified that the autocorrelation measurement allows us characterize both the temporal and spatial properties of high-order LG beams. One reason for obtaining this unanticipated result is that, coincidentally, Eqs. (2) and (3) are mathematically equivalent to the optical autocorrelation functions of LG modes, which can be used to determine the topological charge of both the coherent and partial coherent high-order LG modes without requiring a priori knowledge [42,43]. From another perspective, the results can be interpreted by considering the radial-mode selection rule of LG modes in SHG and the evolution of the Gouy-phase-mediated pattern of superposed LG modes [37, 44, 45]. Regarding the structural evolution of the SHG beams from the near to the far fields, we include additional data in Appendix B. In addition, This interesting phenomenon in the case $\ell = 0$ (i.e., self-imaging of SHG from the near field to the far field) actually exists in a more general sum-frequency generation of two LG modes $LG(\ell_1, p_1) * LG(\ell_2, p_2)$, in which $\ell_1 \times \ell_2 \geq 0$ and $p_1 = p_2$ (see Ref. 37 for further details).

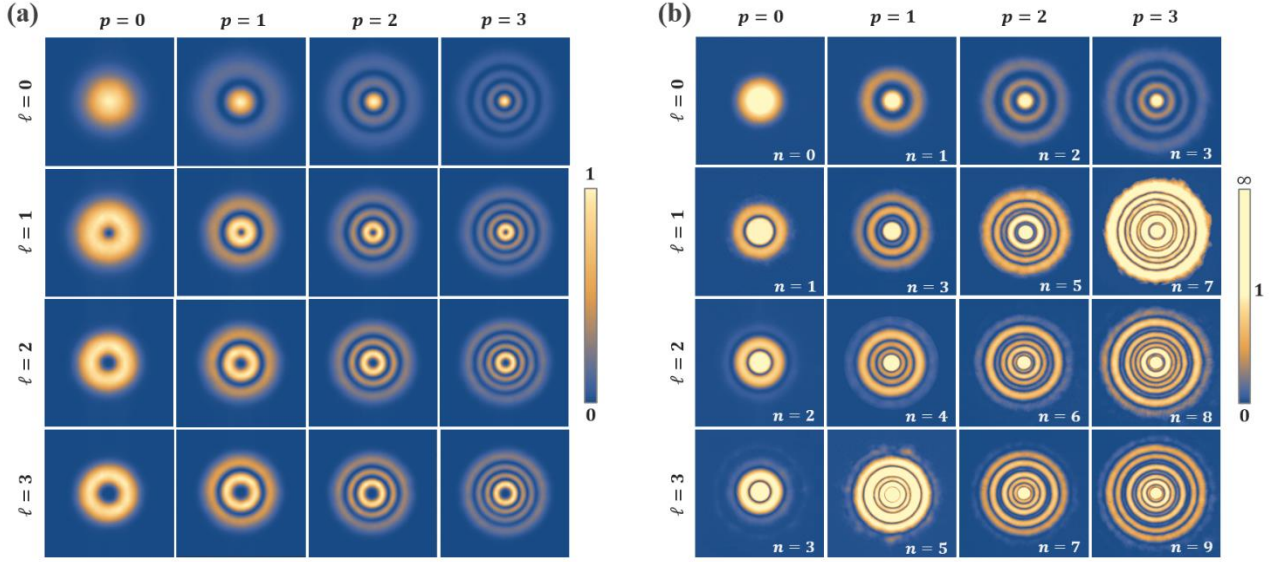

Figure 3. Experimental results: (a) observed intensity profiles of LG modes to be measured, (b) observed intensity profiles of autocorrelation signals (i.e., far-field SHG patterns), corresponding to the theoretical results shown in Figs. 1(a) and 1(b), respectively.

## Conclusion

In summary, we have theoretically proposed and experimentally verified that the autocorrelation measurement, beyond characterizing the temporal structure of light, can also be used to determine the spatial structure of LG modes, including both the azimuthal and radial indices. Specifically, to determine an unknown LG beam, we first determine the radial index by counting the dark-node number $n$ in its original beam profile. Next, if the beam carries nonzero topological charge (i.e., whether a Gaussian mode exists in the centre of the beam profile), we can further determine the modal number $N$ by counting the dark-node number $n$ in the autocorrelation pattern, which gives its azimuthal index via the relation $|\ell| = n - 2p$. These results expand the present nonlinear autocorrelation technique, allowing it to reach its full potential in laser characterization in this era of structured light [46-48].

## Appendix A

Figure S1 show the wavefronts of LG modes and associated autocorrelation signals, corresponding to data in Figs. 1(c) and 1(d). Note that the number of phase jumps shown in Fig. S1(b) determines, for the general case $\ell \neq 0$, how many dark nodes in the autocorrelation patterns.

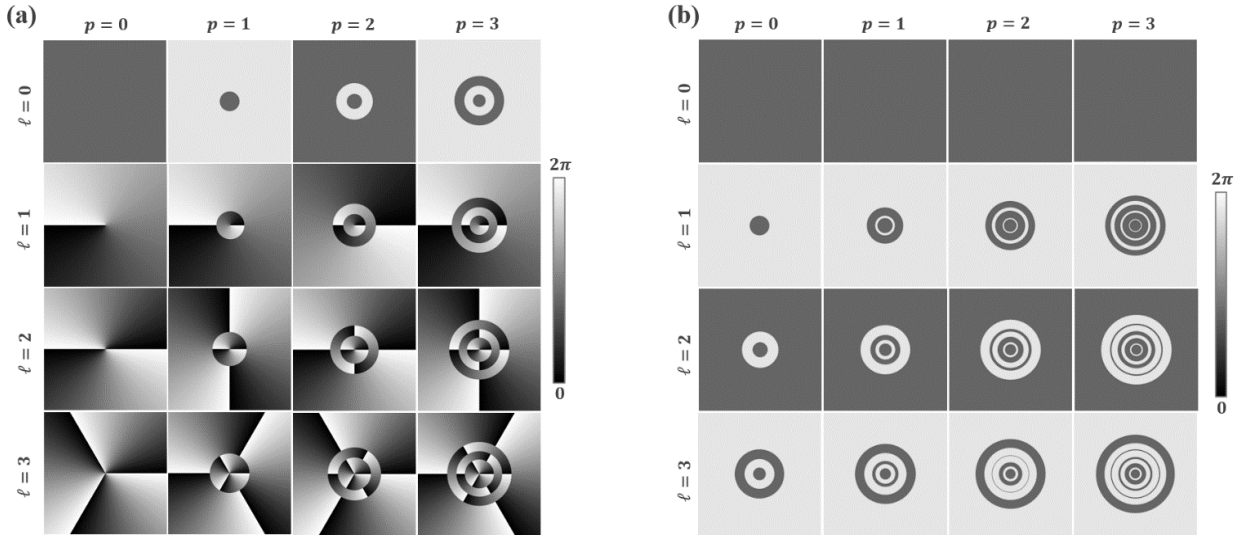

Figure S1. Additional Data: (a) and (b) are theoretical wavefronts correspond to intensity patterns shown in Figs. 1 (c) and 1(d), respectively.

## Appendix B

Here we use two specific examples, corresponding to the two cases shown in Eq. (6), to show the beam-profile evolution of the SHG from the near to the fields, i.e., form Eq. (2) to Eq. (3). For the general example $\ell \neq 0$, Fig. S2(a) shows the structural evolution of spatial autocorrelation of $LG_1^{|2|}$ from the near to the far fields. The corresponding SHG at the generation plane is $E^{2\omega}(r,\varphi,z_0) \propto LG_1^{+2}(r,\varphi,z_0) * LG_1^{-2}(r,\varphi,z_0)$, which can be further expressed as a coherent superposition of LG modes with $\ell = 0$, given by

$$E^{2\omega}(r,\varphi,z_0) = \frac{1}{4}\left[ LG_0^0 - LG_1^0 - LG_3^0 + LG_4^0 \right] \quad (S1)$$

where the component coefficients are obtained by projecting $E^{2\omega}(r,\varphi,z_0)$ onto the corresponding LG bases. The LG components included in Eq. S1 obviously have different modal orders, leading to nonsynchronous Gouy phases accumulated in intramodal phase of the superposition state. As a result, the beam profile of the SHG changes continuously upon propagation, and finally evolves into the structure predicted by Eq. (3) having $n = 2p + |\ell| = 4$ phase jumps.

For the special example $\ell = 0$, Fig. S2(b) shows the autocorrelation of $LG_3^0$, whose SHG field at the generation plane, similarly, can be expressed as

$$\begin{aligned} E^{2\omega}(r,\varphi,z_0) &= \left[ LG_3^0(r,\varphi,z_0) \right]^2 \\ &= \sqrt{25/68}LG_0^0 + \sqrt{9/68}LG_2^0 + \sqrt{9/68}LG_4^0 \\ &\quad + \sqrt{25/68}LG_6^0 \end{aligned} \quad (S2)$$

Coincidentally, the radial indices of LG components (with $\ell = 0$) in this special case are all even numbers. This indicates, for any two components, the difference of their total Gouy phase from the near to the field is 2, 4, or $6\pi$. In consequence, the initial beam profile of the SHG will revive at the Fourier plane, i.e., a self-imaging phenomenon.

## Acknowledgements

This work was supported by the National Natural Science Foundation of China (Grant Nos. 62075050, 11934013, 61975047, 11874102 and 12174047) and the High-Level Talents Project of Heilongjiang Province (Grant No. 2020GSP12).

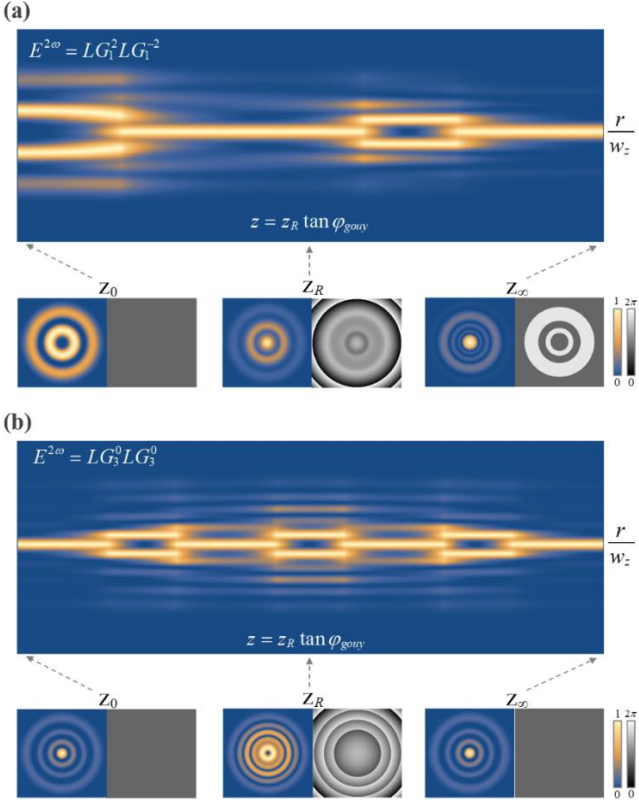

Figure S2. Additional Data: structural evolution of spatial autocorrelation, where examples in (a) and (b) correspond to the general ($\ell \neq 0$) and special ($\ell = 0$) cases, respectively.